\documentclass[11pt,dvips]{article}
\usepackage{epsfig,times}
\usepackage{amssymb,amsmath}

\usepackage{wrapfig}
\usepackage{floatflt}
\makeatletter
\renewcommand{\section}{\@startsection{section}{1}{0in}
	{0.4\baselineskip}{0.1\baselineskip}{\Large\bf}}
\renewcommand{\subsection}{\@startsection{subsection}{2}{0in}
	{0.25\baselineskip}{-\baselineskip}{\large\bf}}
\renewcommand{\subsubsection}{\@startsection{subsubsection}{3}{0in}
	{0.1\baselineskip}{-\baselineskip}{\normalsize\bf}}

\begin{document}
\thispagestyle{myheadings}
\markright{HE.3.2.24}
\begin{center}
{\LARGE\bf Expected Muon Energy Spectra and Zenithal
           Distributions Deep Underwater}
\end{center}
\begin{center}
{\bf A. Misaki$^{1}$, V.A. Naumov$^{2}$, T.S. Sinegovskaya$^{2}$,
     S.I. Sinegovsky$^{2}$, and N. Tahakashi$^{3}$} \\
{\it ${}^{1}$National Graduate Institute for Policy Studies,
             Urawa, 338-8570, Japan \\
     ${}^{2}$Physics Faculty, Irkutsk State University, Irkutsk,
             664003, Russia \\ ${}^{3}$Faculty of Science and
             Technology, Hirosaki University, Hirosaki, 036-8561,
             Japan}
\end{center}

\begin{center}
{\large\bf Abstract \\}
\end{center}
\vspace{-0.5ex}
Energy spectra and zenith-angle distributions of atmospheric muons
are calculated for the depths of operation of large underwater
neutrino telescopes. The estimation of the prompt muon contribution
is performed with three approaches to charm hadroproduction:
recombination quark-parton model, quark-gluon string model, and a
perturbative QCD based model. Calculations show that the larger are
zenith angles and water thickness above the detector, the lower is the
energy  $E_\mu^c$ (``crossing energy'') at which the prompt muon flux
becomes equal to conventional one. For instance, for the depth
of the Baikal Neutrino Telescope and for zenith angle of $78^\circ$
the crossing energy is about 300 TeV, whereas it is only 8 TeV for
the NESTOR depth.  Nevertheless, the muon flux for $E=E_\mu^c$ at
NESTOR depth is in order of magnitude lower in comparison with the
Baikal depth.

\vspace{1ex}

\section{Introduction:}
\label{intro.sec}
The prompt muon (PM) contribution to the atmospheric muon flux
originates from decay of charmed hadrons ($D^\pm$, $D^0$,
$\overline{D}{}^0$, $\Lambda_c^+,\ldots$) that are produced in
collisions of cosmic rays with air nuclei. The problem of charm
hadroproduction, being very important both for particle physics and
high-energy neutrino astronomy, is still remains unsolved.
Modern-day data on high-energy atmospheric muon flux obtained with
many ground-level and underground detectors are too conflicting to be
applicable for a discrimination of charm production models (for a
recent review see Bugaev et al., 1998).
Accuracy of underground measurements is limited due to several
reasons, mainly due to restricted effective volume and uncertainties
in density and chemical composition of the matter overburden.
Therefore, it seems to be interesting to discuss the potentiality for
detecting the PM flux and testing validity of the accepted models for
charm hadroproduction in future high-energy muon experiments with
large underwater neutrino detectors (AMANDA, ANTARES, Baikal NT,
NESTOR). Notice that the atmospheric neutrino induced muon
``background'' becomes negligible for high enough energy threshold.

Current studies of the PM problem apply phenomenological
nonperturbative approaches (see Bugaev et al., 1998) and perturbative
QCD based models (Thunman, Ingelman, \& Gondolo, 1996; Pasquali,
Reno, \& Sarcevic, 1999; Gelmini, Gondolo, \& Varieschi, 1999). The most
 recent pQCD calculations include the next-to-leading
order corrections to the charm production cross sections. Vertical
atmospheric PM flux predicted with pQCD becomes dominant over the
conventional one in the energy range 200 to 1000 TeV; the specific
value of $E_\mu^c$ depends on the QCD model parameters and on the
choice of parton density function set. In present calculations, we
use the quark-gluon string model (QGSM) and recombination
quark-parton model (RQPM) (see Bugaev et al., 1998 and references
therein). We compare our results with ones that follow from the pQCD
based model by Pasquali et al. Notice that the PM flux predicted in
(Gelmini, Gondolo, \& Varieschi, 1999) is essentially larger than the
earlier pQCD prediction (Thunman, Ingelman, \& Gondolo, 1996) and
very close to the results of (Pasquali, Reno, \& Sarcevic, 1999).

\section{Sea-level Muon Fluxes:}
\label{SLMuons.sec}

To calculate the muon spectra and angular distributions at sea level
we apply the atmospheric nuclear cascade model that was described in
detail in (Vall, Naumov, \& Sinegovsky, 1986; Bugaev et al., 1998)
(see also Naumov, Sinegovskaya, \& Sinegovsky, 1998).

\begin{wrapfigure}[26]{r}{8.5cm}
\centering{\mbox{\epsfig{file=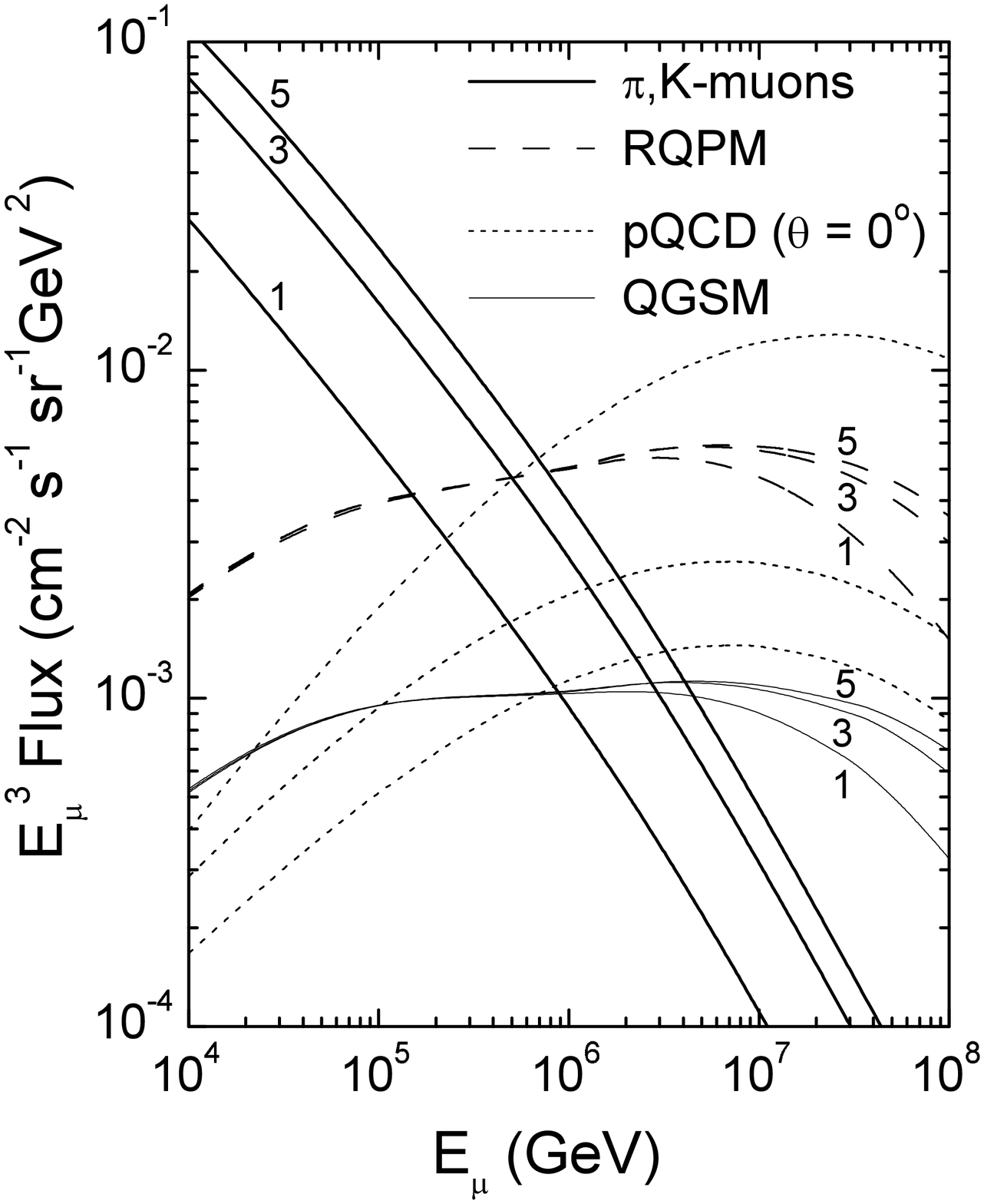,height=10.cm}}
\vskip -18mm
\caption{Sea-level muon fluxes for three zenith angles. The curves
        are for the $\pi,K$-muons (solid) and for the PM
        contributions estimated with the RQPM (dashed), pQCD
        (dotted), and QGSM (thin solid). The numbers shown
        nearby the curves are for values of $\sec(\theta).$
\label{fig-1}}}
\end{wrapfigure}

Differential energy spectra (scaled by factor $E^3$) at sea level
are shown in Fig.~\ref{fig-1} for conventional ($\pi,K$) muons and
for the PM contributions estimated with the RQPM and QGSM for three
directions
corresponding to $\sec(\theta) = 1$, $3$, and $5$, where $\theta$ is
the zenith angle.

In the same figure, we also present the pQCD based results by
Pasquali et al. obtained with three sets of QCD parameters (the
factorization scale $M$, the renormalization scale $\mu$, and the
mass of $c$ quark $m_c$) and with the two sets of parton density
functions \mbox{(STEQ3 and MSRD-)}.  Comparisons with other models
one can find in (Bugaev et al., 1987; Thunman, Ingelman, \& Gondolo,
1996).  The difference among the presented results is caused mainly
by differences between the charm production cross sections.  However,
many assumptions and input parameters (primary cosmic-ray spectrum
and composition, nucleon and light meson production cross sections,
etc.) used in the nuclear-cascade calculations also play a part.
Notice that Pasquali et al. implicitly consider the PM flux to be
isotropic. This is a good approximation for $E_\mu\lesssim10^3$~TeV
and for $\theta\lesssim70^\circ$. But for muon energies and zenith
angles under discussion, the PM flux anisotropy becomes significant
and should be properly taken into account.

As is seen from the figure, the crossing energies $E_\mu^c(\theta)$
for the RQPM case are roughly 140, 480, and 750~TeV for
$\sec(\theta)=1$, $3$, and $5$, respectively, that is close to
the highest pQCD prediction. In the QGSM case, the values of
$E_\mu^c(\theta)$ ($\approx 860$, $2700$, and $4000$~TeV for the
same zenith angles) are fairly close to the lowest
pQCD predictions.

\section{Muon Spectra and Angular Distribution Underwater:}
\label{UWMuons.sec}

The muon energy spectra and zenith-angle distributions deep
underwater are calculated with a semianalytical method (Naumov,
Sinegovsky, \& Bugaev, 1994). By this method one can solve the
problem of muon transport through dense matter for an arbitrary
sea-level muon spectrum and real energy dependence of differential
cross sections for muon-matter interactions. The method is checked
with full Monte Carlo. The calculations of conventional and prompt
muon fluxes underwater at different zenith angles and depths
are performed with all above mentioned charm production models.

\begin{figure}[ht] %%%%%%%%%%%%%%%%%%%%%%%%%%%%%%%%%%%%%%%%%%%%%%%%%%
\centering{\mbox{\epsfig{file=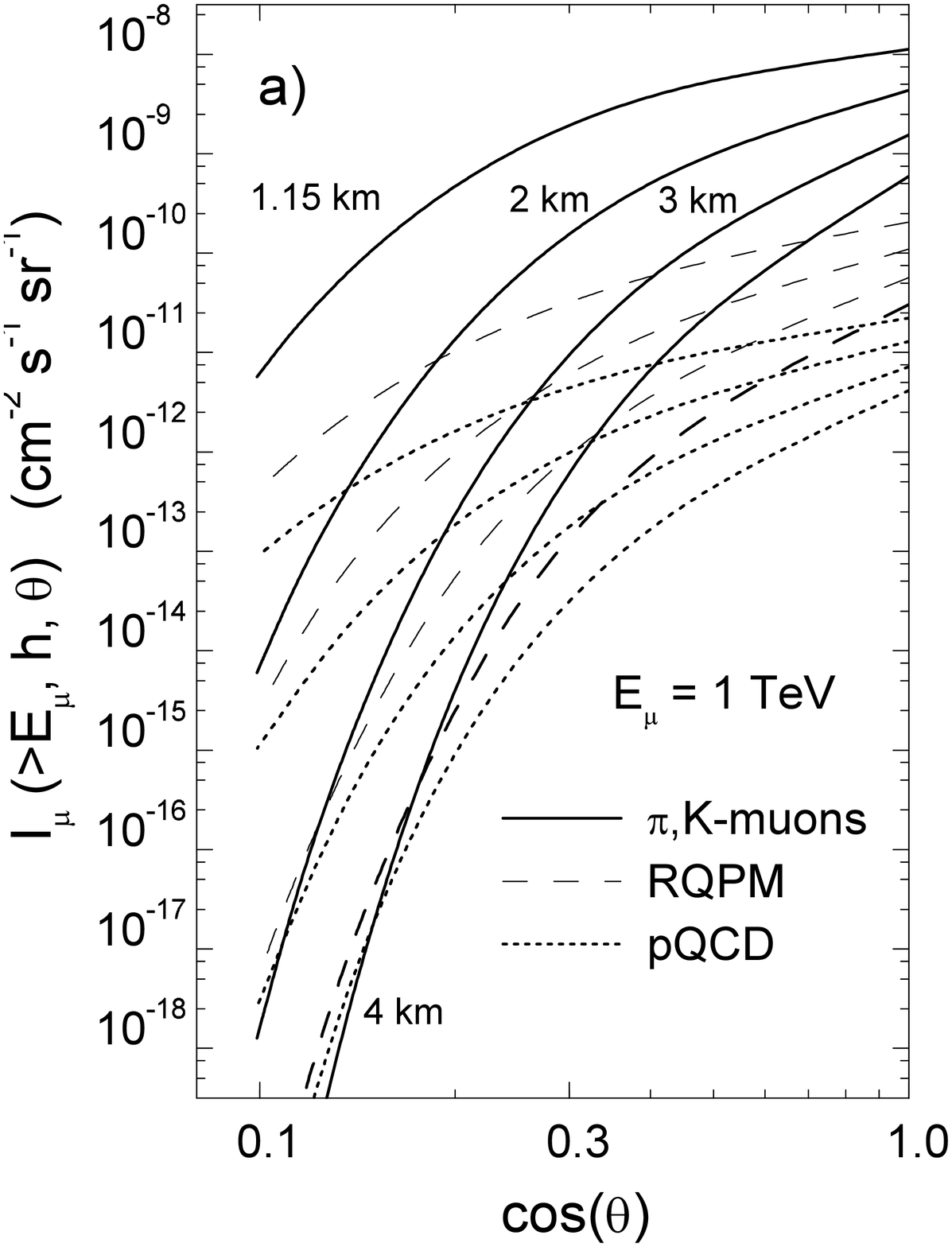,height=10.1cm}}\hspace{5mm}
           \mbox{\epsfig{file=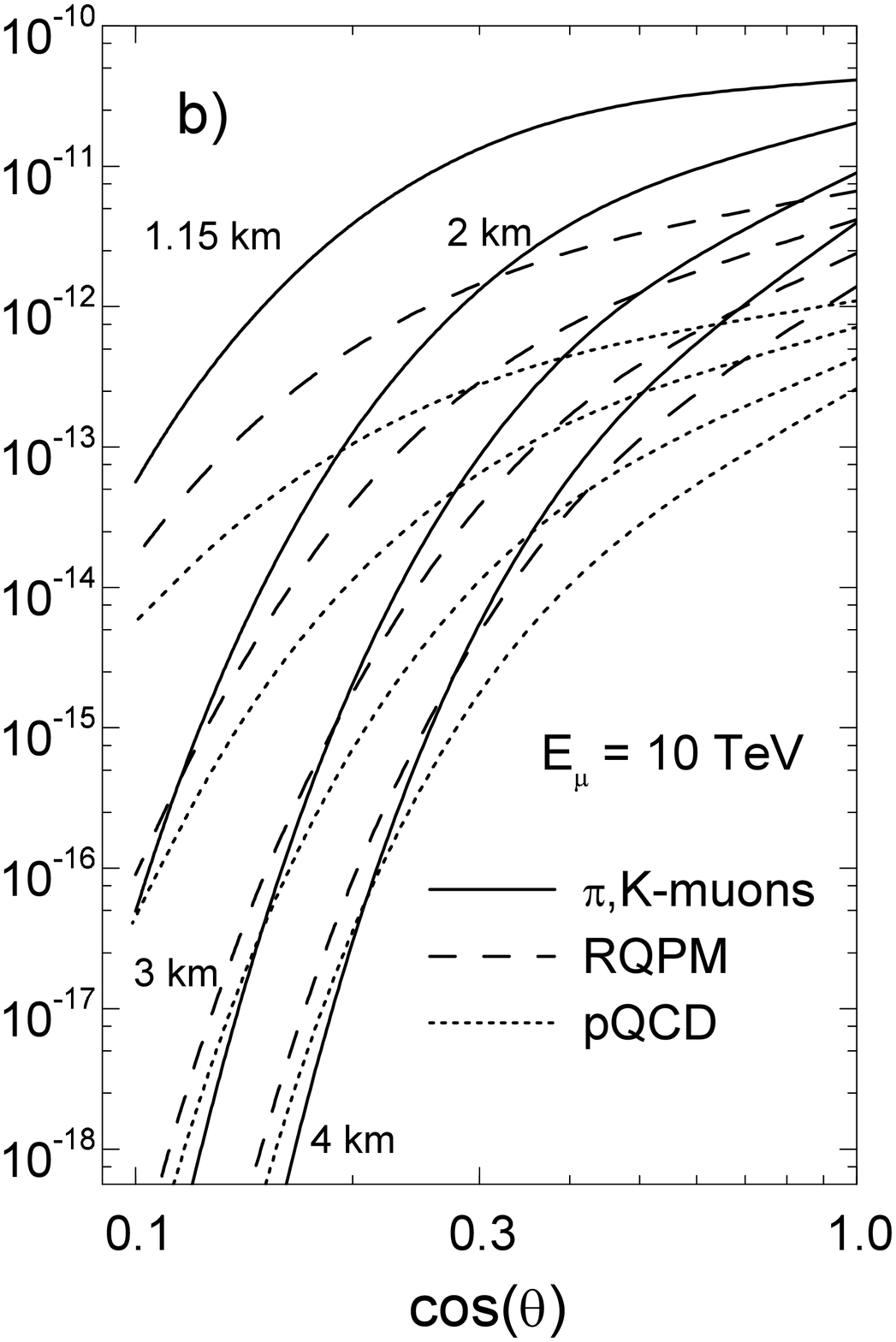,height=10.1cm}}}
\vskip -5mm
\caption{Muon fluxes underwater as a function of the zenith
angle at $E_\mu>1$~TeV (a) and \mbox{$E_\mu>10$~TeV} (b) for depths
$h=1.15$, $2$, $3$, and $4$~km (from top to bottom).
\label{fig-2}}
\vskip 3mm
\centering{\mbox{\epsfig{file=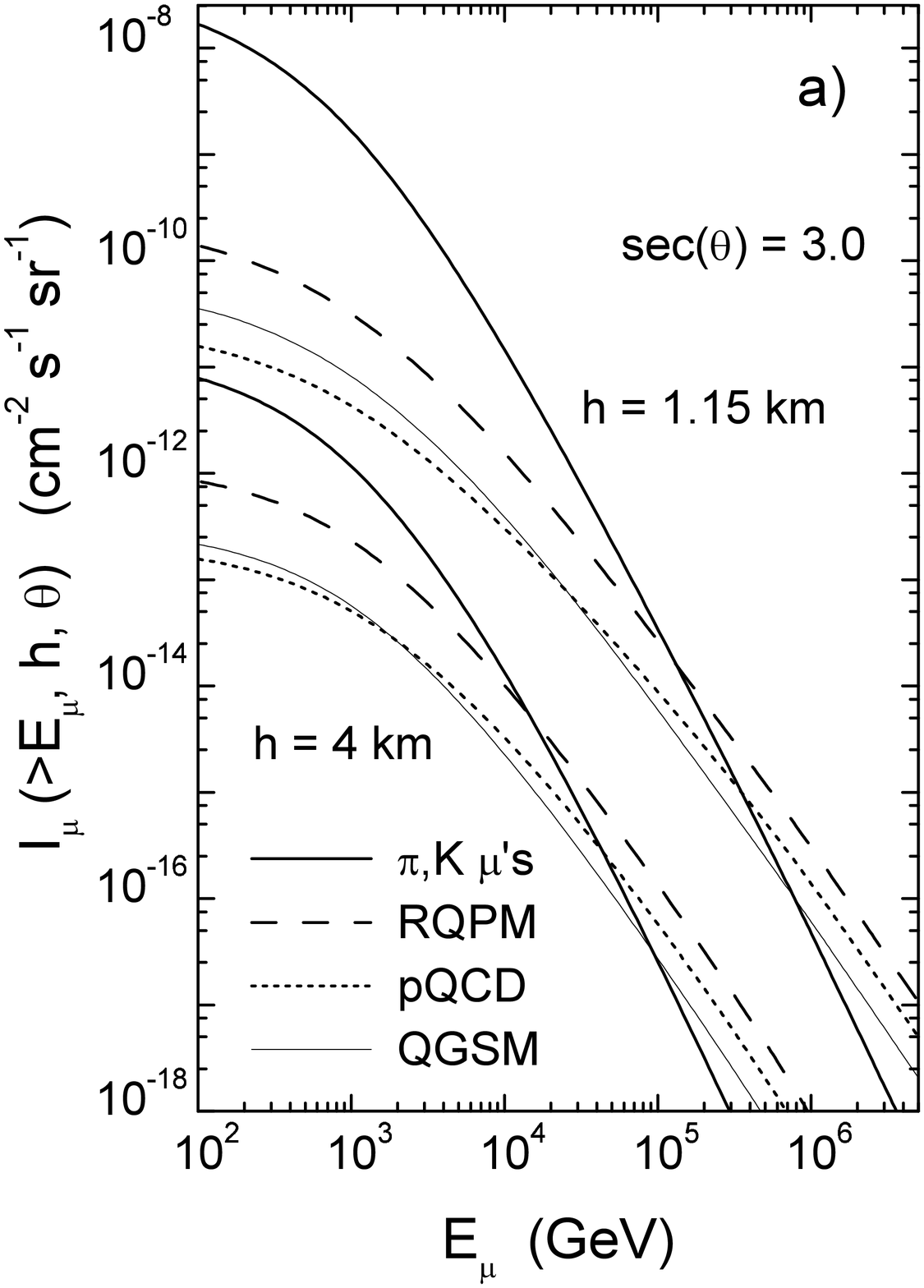,height=10.1cm}}\hspace{5mm}
           \mbox{\epsfig{file=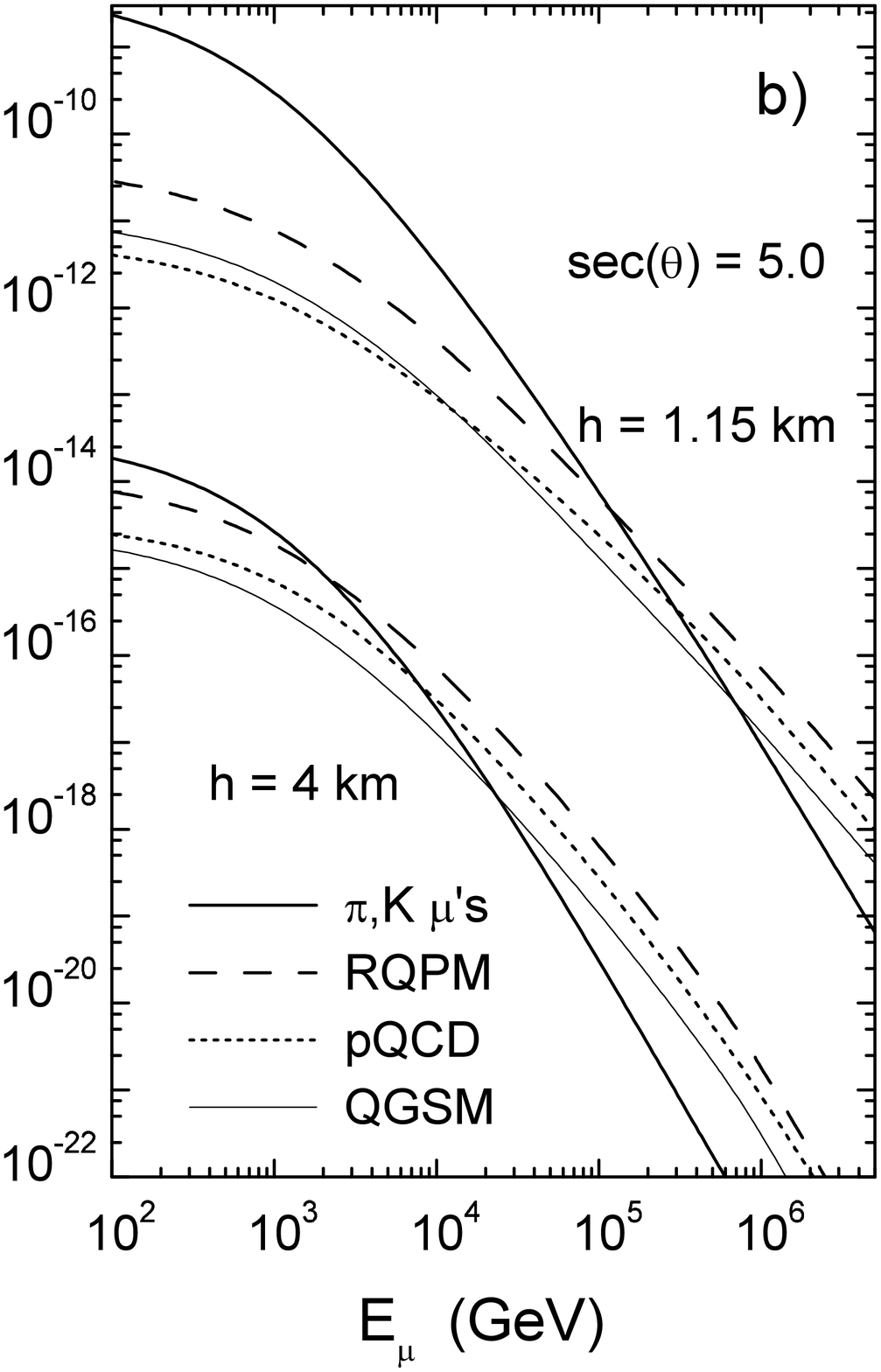,height=10.1cm}}}
\vskip -5mm
\caption{Integral energy spectra of conventional ($\pi,K$)
         and prompt muons underwater for two inclined directions:
         $\sec~(\theta)=3$ (a) and $\sec~(\theta)=5$ (b).
         Four upper curves are for $h=1.15$~km and the rest curves
         are for $h=4$~km.
\label{fig-3}}
\end{figure}

Fig.~\ref{fig-2} shows zenith-angle distributions for the
$\pi,K$ and prompt muons underwater at $E_\mu>1$~TeV and
$E_\mu>10$~TeV for depths $h=1.15$, $2$, $3$, and $4$~km. Here we
present the results obtained with two charm production models,
the RQPM and pQCD. The version of pQCD we use is based on the CTEQ3
parton distributions with $M=2\mu=2m_c$ and $m_c=1.3$~GeV/c$^2$
(this version corresponds to the middle dotted curve in
Fig.~\ref{fig-1}; from here on, we shall call pQCD just this
specific model).

\clearpage
 %%%%%%%%%%%%%%%%%%%%%%%%%%%%%%%%%%%%%%%%%%%%%%%%%%%%%%%%

\clearpage
\noindent
As Fig.~\ref{fig-2} suggests, for not-too-deep water
($h<1.5-2$~km) there is no intersection between the curves which
represent the conventional and prompt muon fluxes at
$\theta\lesssim85^\circ$ and $E_\mu\lesssim10$~TeV. The intersection
point shifts to smaller zenith angles with increasing depth.

The absolute value of the muon flux in this point drastically depends
on the charm production model. This is a promising fact which is able
to help in the establishment of experimental constraints to the charm
hadroproduction models from the measuring the muon zenith-angle
distribution for high enough detection threshold.

Fig.~\ref{fig-3} shows integral energy spectra of muons for
$h=1.15$~km (the Baikal NT depth) and $h=4$~km (the NESTOR depth) and
for $\sec(\theta)=3$ ($\theta\simeq70.5^\circ$) and $\sec(\theta)=5$
($\theta\simeq78.5^\circ$). The predictions of three charm production
models are presented. The crossing energies for the NESTOR depth are
essentially lower in comparison with ones for the Baikal depth (a
factor of about 8 for $\sec(\theta)=3$ and of 35 to 60 for
$\sec(\theta)=5$). In particular, for $\sec(\theta)=5$,
$E_\mu^{c\,\mathrm{(pQCD)}}\approx8$~TeV for NESTOR while
$E_\mu^{c\,\mathrm{(pQCD)}}\approx300$~TeV for Baikal. Nevertheless,
above the crossing energies, the muon flux is almost in order of
magnitude higher for the Baikal depth.

\section{Conclusions:}
\label{Concl.sec}

Energy spectra and zenith-angle distributions of atmospheric muons
have been calculated for the depths 1 to 4~km that correspond the
depths of operation of large underwater neutrino telescopes. The
estimation of the sea-level prompt muon contribution performed with
RQPM, QGSM and pQCD shows that the energy, at which the prompt muon
flux becomes equal to conventional one (``crossing energy''), spreads
within a wide energy range 140 to 4000~TeV.

For large zenith angles, the crossing energies (and hence the
necessary detection thresholds) are in order of magnitude larger for
the operation depth of the Baikal detector in comparison with ones
for the NESTOR depth. Despite of this fact, the Baikal depth proves
to be more suitable for the problem under discussion, compared to
the NESTOR one (all other things being equal) considering that
the absolute muon intensity for $E>E_\mu^c$ at the NESTOR
depth is almost in order of magnitude lower. More generally,
comparatively small depths ($1-2$~km) and not-too-large zenith angles
($\theta\lesssim80^\circ$) have certain advantages for future
underwater experiments with prompt muons.

\section*{Acknowledgements:}

The work of V.\,N., T.\,S., and S.\,S. is partially supported by the
Ministry of General and Professional Education of Russian Federation
under Grant No.~728 within the framework of scientific program
``Universities of Russia -- Basic Researches''.

\vspace{1ex}
\begin{center}{\Large\bf References}\end{center}
 Bugaev, E.V. et al. 1998, Phys. Rev. D58, 054001 \\
 Bugaev, E.V. et al. 1989, Nuovo Cim. C12, 41 \\
 Gelmini G., Gondolo P., \& Varieschi G. 1999, hep-ph/9904457 \\
 Naumov, V.A., Sinegovsky, S.I., \& Bugaev, E.V. 1994, Yad. Fiz. 57,
         439 [Phys. At. Nucl. 57, 412]  \\
 Naumov, V.A., Sinegovskaya, T.S., \& Sinegovsky, S.I. 1998, Nuovo
         Cim. 111A, 129 \\
 Pasquali, L., Reno, M.H., \& Sarcevic, I. 1999, Phys. Rev. D59,
         034020 \\
 Thunman M., Ingelman G., \& Gondolo P. 1996, Astrop. Phys. 5,
         309 \\
 Vall, A.N., Naumov, V.A., \& Sinegovsky, S.I. 1986, Yad. Fiz. 44,
         1240 [Sov. J. Nucl. Phys. 44, 806]
\end{document}